\documentclass[reprint,twocolumn,amsmath,amssymb,pra]{revtex4}
\usepackage{graphicx}

\newcommand{\bra}[1]{\langle#1|}
\newcommand{\ket}[1]{|#1\rangle}

\begin{document}

\title{Frequency and phase relations of entangled photons observed by a two-photon interference experiment.}

\author{Joakim Bergli}%
 \email{joakim.bergli@fys.uio.no}
\author{Guillaume Adenier}
\author{Andreas P. Th\"{o}rn}%
\author{Arnt Inge Vistnes}%

\affiliation{%
 Department of Physics, University of Oslo, Sem S{\ae}lands vei 24, 0316 Oslo, Norway
}

\date{\today}

\begin{abstract}
An entangled photon experiment has been performed with a large
variation of the temperature of the non-linear crystal generating the
entangled pair by spontaneous downconversion. The photon pairs are
separated by a nonpolarizing beamsplitter, and the polarization modes are
mixed by half wave plates. The correlation function of the coincidences is
studied as a function of the temperature. In the
presence of a narrow interference filter we observe that the correlation changes
between -1 and +1 about seven times within a temperature interval of
about 30 degrees C. We show that the common simplified single-mode pair representation of entangled photons is insufficient to describe the results, but that the biphoton
description that includes frequency and phase details gives close to perfect fit with experimental data for
two different choices of interference filters. We explain the main
ideas of the underlying physics, and give an interpretation of the two-photon amplitude which provides an intuitive understanding of the effect of changing the temperature and inserting interference filters. 
\end{abstract}

\maketitle

\section{Introduction}

One of the central experiments in quantum optics is the Hong-Ou-Mandel
(HOM) two-photon interference experiment~\cite{HOM}. In the original
version, two entangled photons with the same polarization were
generated by downconversion, and directed on a beamsplitter from
opposite sides. With the timing adjusted so that the two photons came
to the beamsplitter at the same time, it was observed that the two
photons always exited in the same direction. This could then be
interpreted as the interference of two independent photons, where the
coalescence of the two photons was seen as a consequence of the
bosonic nature of photons. However, similar experiments were soon
devised where the two photons reached the beamsplitter at different
times~\cite{pittman}, where they were distinguishable by different
polarizations~\cite{Kim2005}, or not even had to meet in a
beamsplitter~\cite{sergienko1995}. It became clear that the proper way
of interpreting these experiments had to be in terms of interference between
two different two-photon states.  For a more complete discussion of
the different experiments and review of the literature, see
Ref~\cite{Shih}.

The state of the field after the downconversion process is well
known~\cite{Rubin94,TSuhara,TSuhara2007,Martin08,Martin10,KuklewiczPhD,Sciarrino}
and consists of a superposition of many frequency modes, a brief
derivation is given below. However, in many situations (see for
instance
\cite{Rubin94,TSuhara,TSuhara2007,Martin08,Martin10,KuklewiczPhD,Sciarrino})
 one can simplify the description to a single
frequency mode and two orthogonal polarizations. Nevertheless, there are cases where this is insufficient, and the spread in frequencies is important for the understanding of experiments~\cite{fedrizzi,brida}.  As demonstrated  by Fedrizzi {\it et al.}~\cite{fedrizzi} one can  reveal what they call
``hidden entanglement''. That is, that the state is indeed frequency
entangled.  This is done by changing the temperature of the nonlinear crystal where the
photon downconversion takes place. Each such
crystal has an optimal working temperature where the downconversion
process is most efficient, and where the energies of the two outgoing
photons are equal (on average). At other temperatures, the two photons
will have different average energies. We present a different
experimental setup which allows essentially all possible correlation
measurements to be performed in a two-photon interference setup. With
this we study how the correlations between the two photons vary with
the temperature of the nonlinear crystal is varied. We can then
observe the frequency entanglement as demonstrated in
Ref~\cite{fedrizzi} but in a much more dramatic way. Using an interference filter, we restrict the
number of frequency modes available, and investigate to what extent we can recover
the predictions of the simplified, single frequency mode
description. We find that at the optimal temperature of the nonlinear
crystal, the single frequency mode is always aplicable, and we
identify ``difficult'' temperatures, where extremely narrow filters
would have to be used in order to restore the single-frequency picture
(or alternatively, where the frequency entanglement produces
pronounced effects, even if only a narrow range of frequencies is
involved). We provide a theoretical description that reproduces all
experimental results, and we show how it can be used to understand why
frequency entanglement is irrelevant at the optimal temperature, and
why it becomes relevant at other temperatures. A preliminary report of
our experiments is available in \cite{adenier}, where more experimental
results are given. Here we reproduce only those results that are
directly relevant to our discussion, and provide a detailed
theoretical analysis of the results.

\section{Experimental setup}\label{Expsetup}

\begin{figure}
\includegraphics[width=8.75cm]{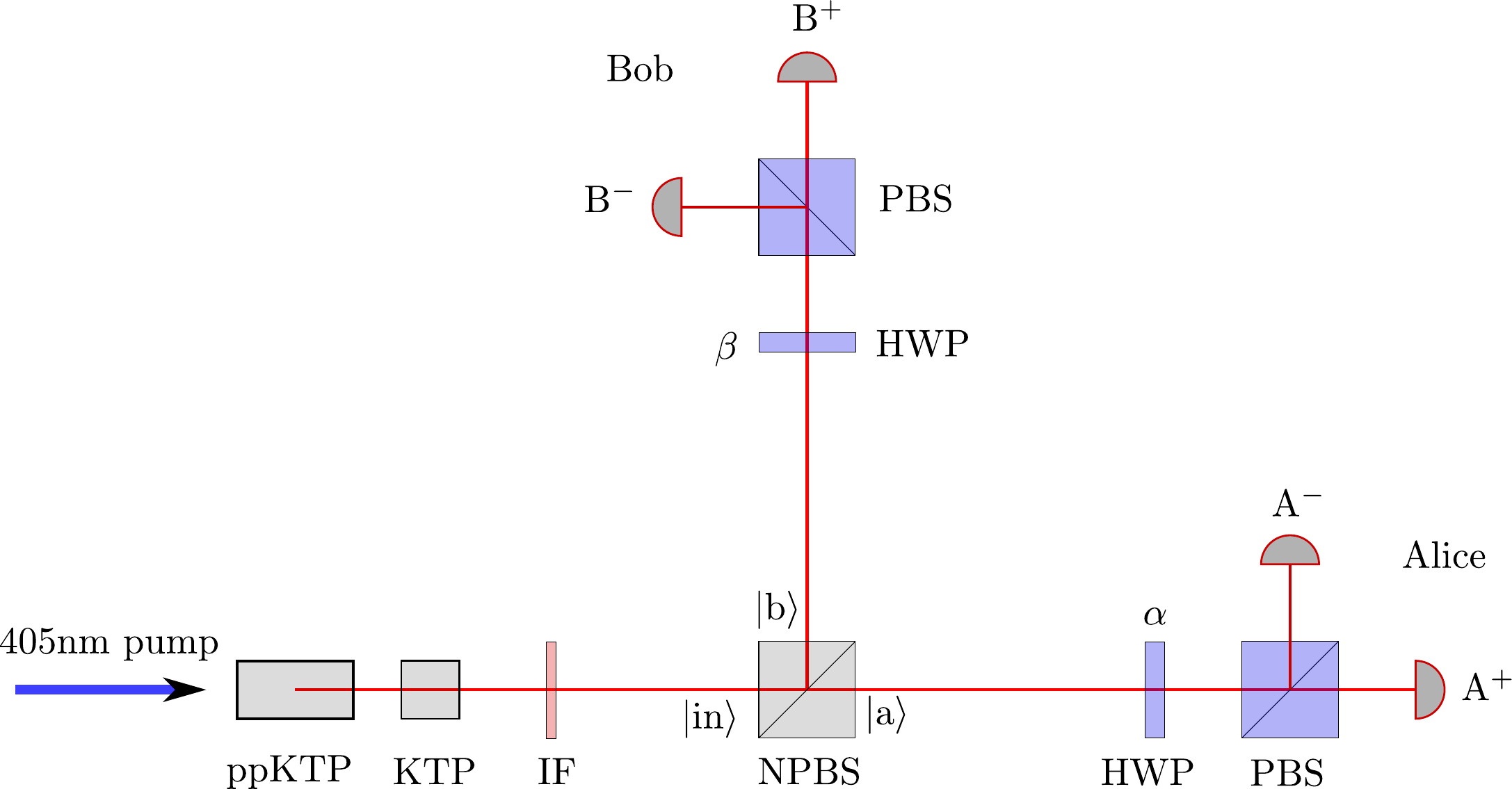}
\caption{\label{RTfig} Experimental setup. The pairs of orthogonally
  polarized photons are down converted from the 405 nm pump in the
  ppKTP crystal. They are time compensated (in a KTP crystal),
  frequency filtered (IF), and spatially filtered (iris) before being
  dispatched to Alice and Bob by a non-polarizing beam splitter
  (NPBS). At each measuring station, a half-wave plate (HWP) rotates
  the polarization of the photons, and a PBS projects them to a fixed
  basis. The output of the PBS feed four detectors, $\mathrm{A}^+$ and
  $\mathrm{A}^-$ for Alice, $\mathrm{B}^+$ and $\mathrm{B}^-$ for
  Bob.}
\end{figure}

Our source of polarization-entangled photons (see Fig.~\ref{RTfig}) is directly inspired by
the setup implemented by Kuklewicz et al.\cite{Kuklewicz}. The pairs
of photons are obtained by type-II spontaneous parametric down
conversion in a periodically-poled crystal of
potassium-titanyl-phosphate (ppKTP), under quasi-phase matching (QPM)
condition. The pump, a continuous-wave laser at 405 nm, and the down
converted photons are all collinear. The temperature of the ppKTP
crystal is controlled by a thermoelectric Peltier temperature
controller.

After filtering out the pump with a series of dichroic and interference filters (IF) (in particular one with a bandwidth of 1 nm), the collinear down converted photons are dispatched by a 50/50 non-polarizing beam splitter (NPBS) to two polarization analyzers. We label the polarization analyzer in the transmitted beam as ``Alice'', and the one in the reflected beam as ``Bob''. Each consists of a half-wave plate that rotates the polarization of the field, with an angle $\alpha$ for Alice and $\beta$ for Bob, followed by a polarizing beam splitter (PBS) that projects it in a fixed basis $\{|H\rangle,|V\rangle\}$.

The outputs of each PBS are feeding two detectors, labeled
$\mathrm{A}^+$ and $\mathrm{A}^-$ for Alice---respectively located at
the transmitted and reflected output of Alice's PBS---$\mathrm{B}^+$
and $\mathrm{B}^-$ for Bob---respectively
located at the transmitted and reflected output of Bob's PBS. The
detectors are four avalanche photodiodes SPCM-AQRH-16, from
Perkin-Elmer, with a detection efficiency specified at 60\% by the
manufacturer with 25 dark counts per seconds. The detection events
(clicks) are time-tagged with nominal picosecond precision by a
Multichannel Picosecond Event Timer (Hydraharp 400, from Picoquant),
and saved to disk for on-the-fly analysis (when the data flow is not
too important), and also for subsequent analysis.

The flexibility of this acquisition setup based on the recording of
the detection time of the photons is of course largely inspired by the
landmark Innsbruck experiment performed by Weihs et
al. \cite{Weihs}.

Unless specified otherwise, the acquisition duration for each measured
point lasted precisely one second, so that the number of counts
recorded during this interval can be used as an estimate of the rate
of the corresponding counts.

All detected events are recorded with the name of the detector that
fired and the time associated with the detection event. No detected
events are discarded, so that the coincidence analysis can be
performed after-the-fact with adjustable parameters (size of the
coincidence window and timing-offset). Having a
complete record of all the detected events allows us to be thorough in
the coincidence analysis: we can naturally measure several type of
rates. The rate of single counts of course, but also the rate of
coincidences, the rate of double-counts, or even the triple-counts.

A rate of coincidences denotes the number of times one of Alice's
detectors triggers within the same time window (of arbitrary width) as
one of Bob's detectors, during an acquisition of 1 second. The four
possible rates of coincidences between Alice's and Bob's detectors are
denoted $R_\mathrm{ab}^{++}$, $R_\mathrm{ab}^{+-}$,
$R_\mathrm{ab}^{-+}$ and $R_\mathrm{ab}^{--}$, where the first
superscript index indicates which of Alice's detector is considered,
and the second which of Bob's detector is.

A rate of double-counts denotes the number of time the two detectors
located on the same side (Alice's side or Bob's side) are both triggered
within the same time window, during an acquisition of 1 second. The
rate of double-counts measured by Alice are denoted
$R^{\pm}_\mathrm{aa}$; those measured by Bob are denoted
$R^{\pm}_\mathrm{bb}$.

\section{Experimental results}\label{exp}

The downconversion process is most efficient at an optimal temperature
of the ppKTP crystal (in our case $35.1^{\circ}\mathrm{C}$), for which
the number of produced polarization entangled pairs is maximal. We
study the properties of our entangled photons when the temperature of
the ppKTP crystal is brought away from this optimal temperature.

\begin{figure}
\includegraphics[width=8.75cm]{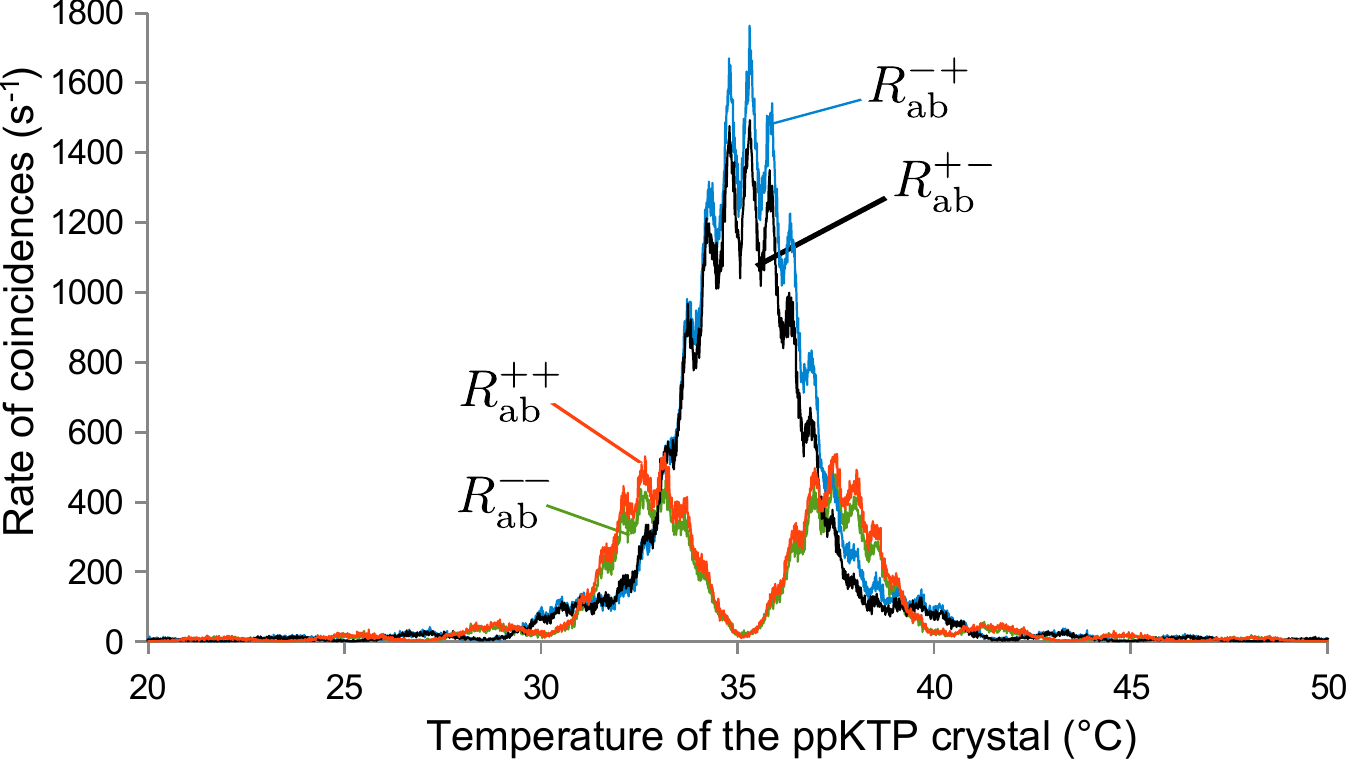}
\caption{\label{CoincidencesNonOptTemp} Rates of coincidences in the diagonal bases ($\alpha=\beta=\pi/8$) as a function of the temperature of the ppKTP crystal. The rate of coincidences drops quickly away from the optimal temperature because the distribution of the wavelengths of the idler and signal are no longer centered on the 810 nm of our narrow bandwidth interference filter. The rapid oscillations in the coincidence rates arise from the reflections of the pump beam from the ends of the ppKTP crystal, which gives alternatingly constructive and destructive interference between the direct and reflected beams, and an effective oscillation in the pump amplitude. These oscillations are not interesting for us, and they are absent from the correlations as decribed by Eq. \eqref{correlation} since an overall intensity is normalized out in calculations of correlations.}
\end{figure}

In Fig. \ref{CoincidencesNonOptTemp} we show the coincidence rates
as functions of the ppKTP temperature for the case where the axes of
the HWPs are set to $\alpha=\beta=\pi/8$, which means that both
horizontally and vertically polarized photons are rotated into equal
superpositions of horizontal and vertical polarization, as described
by Eq. \eqref{HWPtransfo} below.  We refer to this as a measurement in
the diagonal bases.

The first noticeable effect of bringing the temperature of the ppKTP
crystal away from the optimal temperature is that the rate of detected
singles and coincidences quickly drops, as can be seen for the
coincidences in Fig.~\ref{CoincidencesNonOptTemp}. The reason for this
behavior is that the spectrum of the down-converted photons depends on
the temperature of the ppKTP crystal. Indeed, all the terms in the
quasi-phase matching condition of Eq.~(\ref{kSPDC}) depend on the
temperature of the ppKTP crystal \cite{Emanueli,Hamel}. At the optimal
temperature, both outgoing beams have the same frequency,
corresponding to a wavelength $\lambda=810$ nm, exactly twice that of
the pump beam. As the temperature deviates from the optimal, the
center frequencies of the two beams start to differ, while keeping the
sum constant in accordance with the energy conservation condition,
Eq.~\eqref{omegaSPDC} below. Now, because of the interference filter with a
narrow bandwidth of 1 nm in our setup (see Fig.~\ref{RTfig}), we are
nevertheless selecting those pairs of photons that happen to have the
\emph{same} wavelength of 810 nm. As the temperature is changed away
from the optimum temperature, the pairs of photons that match this
strict wavelength criterion are less and less frequent, so that the
number of coincidences drops quickly.

Even though the number of pairs that passes through diminishes away
from the optimal temperature, we can still measure their coincidences,
and their correlation. The correlation 
\begin{equation}\label{correlation}
E_\mathrm{ab} (\alpha,\beta)=
\frac{R^{++}_\mathrm{ab}-R^{+-}_\mathrm{ab}-R^{-+}_\mathrm{ab}+R^{--}_\mathrm{ab}}
     {R^{++}_\mathrm{ab}+R^{+-}_\mathrm{ab}+R^{-+}_\mathrm{ab}+R^{--}_\mathrm{ab}},
\end{equation}
is a linear combination of the coincidence rates normalized by the sum
of coincidences rates. It is therefore quite insensitive to
fluctuations in the total rate of detected pairs. Reducing the number
of detected pairs decreases the statistical accuracy of the measured
correlation, but it does not change this correlation per se. It can be
compensated simply by increasing the power of the pump accordingly,
which we have done in some of the experimental runs reported below
when the rate of detected pairs was too low.

\begin{figure}
\includegraphics[width=\linewidth]{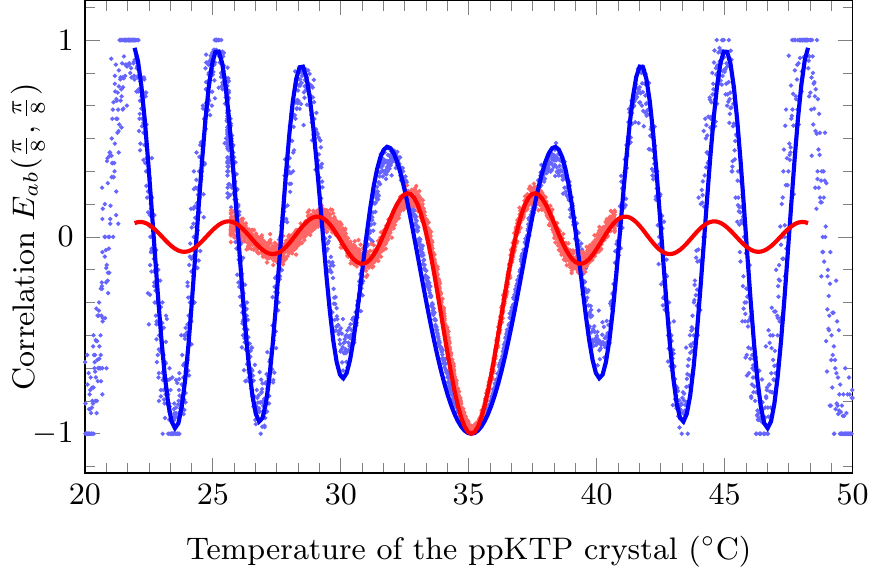}
\caption{\label{CorrelationNonOptTemp} Correlation in the diagonal
  bases ($\alpha=\beta=\pi/8$) as a function of temperature of the
  ppKTP crystal. Blue points are data acquired vith a 1 nm IF while
  red points are acquired with a 10 nm IF. The corresponding curves
  are theoretical predictions. Although the rate of coincidences drops
  quickly away from the optimum temperature of $35.1^\circ$C, the
  correlation does not in any way disappear, but oscillates with
  increasing amplitudes. The occurrences of positive correlation close
  to 1, indicating that the photons share the same polarization in the
  diagonal basis, is particularly worth investigating, given that we
  are operating in type II spontaneous parametric downconversion
  conditions, i.e., with orthogonally polarized photons.}
\end{figure}

Quite generally, a direct way to assess and fine-tune the quality of
the produced polarization-entanglement is to measure the correlation
when Alice and Bob have their settings set at diagonal in polarization
space, which corresponds to $\alpha=\beta=\pi/8$ for the half-wave
plates located in front of their respective PBS. Indeed, it is in the
diagonal bases that the visibility of the correlation is naturally the
lowest, and any departure from the optimal conditions reduces the
absolute value of this correlation, whereas in the horizontal or
vertical basis it is much less sensitive to imperfections.
We have therefore measured the correlation in the diagonal bases while
varying the temperature of the ppKTP crystal. The result is displayed
in Fig.~\ref{CorrelationNonOptTemp} (blue points).

At optimum temperature ($35.1^{\circ}\mathrm{C}$), the correlation in
the diagonal bases $\alpha=\beta=\pi/8$ is close to $-1$.  For small
temperature variation of the ppKTP crystal away from the optimal
temperature, the absolute value of the correlation in the diagonal
bases decreases, which could be tempting to interpret as caused by a
loss of indistinguishability between the photons, as  the centers
of the spectral distributions of the signal and idler photons start to
differ more and more. However, the surprising feature revealed in
Fig.~\ref{CorrelationNonOptTemp} is that when departing further away
from the optimum temperature, the correlation does not remain close to
zero---as would be expected from distinguishable H and V photons
observed in the diagonal bases---but oscillates instead with
increasing amplitude, until the correlation reaches again absolute
values close to unity.

Of particular interest are the temperature values for which the
correlation becomes positive and close to 1, which happens below the
optimum temperature at 28.6$^{\circ}\mathrm{C}$,
25.0$^{\circ}\mathrm{C}$, and $21.8^{\circ}\mathrm{C}$. Indeed, it
then means that the photons measured in the diagonal bases
($\alpha=\beta=\pi/8$) share the same polarization, which is
surprising given that the downconverted photons are of type-II, that
is, orthogonally polarized. We have observed that the orthogonality
can actually still be seen quite clearly when Alice's and Bob's fix
the orientation of their half-wave plates at $\alpha=\beta=0$
instead. The correlation is then very close to $-1$.

It should be noted that these experimental features depend strongly on
the use of the 1 nm bandwidth interference filter. When we instead used
 an interference filter with 10 nm bandwidth
(Fig.~\ref{CorrelationNonOptTemp}, red points), the minima and maxima
were shifted in temperature, and the amplitudes of the oscillations
were greatly reduced, except at the optimum temperature where the
correlation was still quite close to $-1$. Indeed, the results are
then very similar to those observed in \cite{fedrizzi} without any
filter. Even if the experiment of Fedrizzi et al. is different in the setup relative to ours (separating
the two polarization states and recombining them in a beam splitter as
in the original HOM experiment), it is closely related to our
experiment in the analysis. In \cite{fedrizzi}, positive correlation
indicates that both photons exit the beamsplitter at different output
ports. That is, it indicates photon antibunching, which implies a
state that is antisymmetric in polarization.

The situation is different in our case because the photons are
collinear and incident on the \emph{same} input port
$|\mathrm{in}\rangle$ of the beam-splitter. The spatial modes of the
idler and signal photons are also not distinguished in our
experimental setup, so that it is a priori the same for both photons,
and therefore symmetric by particle exchange. The compensation of the
antisymmetry of the polarization degree of freedom is then not
immediately obvious. As discussed in Ref.~\cite{fedrizzi}, one
possible explanation comes from the fact that the photon state is
frequency entangled. That is, each outgoing beam has a certain
frequency spread, but the frequencies of the photons are not
independently distributed over the frequency width of the
beam. Rather, if one photon has a frequency higher than the center
frequency of that beam, the other photon will have a frequency below
the center frequency of the other beam, so that the energy
conservation condition \eqref{omegaSPDC} is satisfied for each photon
pair. As will be seen from the analysis below, our experiment, while
different in setup, is logically completely equivalent to that in
Ref.~\cite{fedrizzi}, except that we have used an additional
interference filter to reduce the frequency spread of the two
beams. Comparing the results with and without this filter, we make two
interesting and initially surprising observations:

\begin{enumerate}
\item
The apperance of positive correlation, indicating antisymmetry of the
polarization state, is explained by the presence of a corresponding
antisymmetry in the frequencies, rendering the full state symmetric as
expected for photons. That is, the observations are only explained if
there is a certain frequency spread of the downconverted
beams. Inserting a filter which limits this spread, one would expect
the observed positive correlation to be reduced, whereas we observe
the opposite. The correlation increases, reaching almost the maximum
of +1.

\item
The perfect negative correlation at the optimal temperature is insensitive to the filter width. It appears to be the same for all possible frequency spreads. 

\end{enumerate}
In the following we will address these issues, and give detailed explanations of the observations.

\section{Theoretical description}

\subsection{Spontaneous parametric down conversion}
The theoretical description of polarization entanglement from
collinear type-II spontaneous parametric down conversion is
well known (see for instance
\cite{Rubin94,TSuhara,TSuhara2007,Martin08,Martin10,KuklewiczPhD,Sciarrino}). In
a nonlinear crystal, the interaction Hamiltonian is \cite{Rubin94}
\begin{equation}\label{Hamilton }
    \hat{H}=\epsilon_0\int_V \mathop{dr^3} \chi^{(2)}\hat{E}^+_\mathrm{p}\hat{E}^-_\mathrm{1}\hat{E}^-_\mathrm{2} \;+ \; h.c,
\end{equation}
where the index $p$ stands for \emph{pump}, while 1 and 2 refer to the
two output modes (\emph{signal} and \emph{idler}). A photon from the
pump can be spontaneously down-converted to two daughter photons.  In
a periodically poled crystal, this process must fulfill the quasi
phase-matching (QPM) conditions, for the angular frequencies
\begin{equation}\label{omegaSPDC}
    \omega_\mathrm{p}=\omega_\mathrm{1}+\omega_\mathrm{2},
\end{equation}
and for the wave numbers
\begin{equation}\label{kSPDC}
    k_\mathrm{p}=k_\mathrm{1}+k_\mathrm{2}+\frac{2\pi}{\Lambda},
\end{equation}
where $\Lambda$ is the period of the poling in the ppKTP crystal. For
a crystal of finite length, Eq~\eqref{kSPDC} only has to be satisfied
approximately, which results in a certain finite width of the spectral
peak for the downconverted photons.

For collinear spontaneous parametric downconversion confined to a single spatial mode, the calculation
to first order perturbation theory for the quantum state of the pairs
of down-converted photons at the output of the nonlinear crystal is
\cite{Teich02,Nam03}
\begin{equation}\label{SPDC}
  \ket{\psi} = W\int d\nu
  f(\nu)
  a_{in,H}^\dag(\omega_1) a_{in,V}^\dag(\omega_2)\ket{0}
\end{equation}
where $a_{in,H}^\dag(\omega)$ and $a_{in,V}^\dag(\omega)$ are the
creation operators for signal and idler photons in the input port
$|\mathrm{in}\rangle$ of the beam-splitter with horizontal and
vertical polarization in frequency mode $\omega$, and where the
integral is taken from $-\infty$ to $+\infty$.  The function
\begin{equation}\label{fnu}
 f(\nu) =  \int_{-L}^{0}dz e^{i Dz\nu} 
\end{equation}
describes the spectral distribution of the downconverted photons. Here $L$ is
the length of the ppKTP crystal, and we have assumed that it is
oriented along the $z$-direction from $z=-L$ to $z=0$. We have defined
\begin{equation}
  D =  \frac{1}{c_2}-\frac{1}{c_1}
\end{equation}
where $c_1$ and $c_2$ are the light speeds for the two output beams
within the crystal. We let $\omega_1^0$ and $\omega_2^0$ be the output
frequencies that satisfy both Eqs. \eqref{omegaSPDC} and
\eqref{kSPDC}, which means that they can be interpreted as the central frequencies
of the spectral distributions of the two photon beams. The frequency
deviation $\nu$ is then given by
\begin{equation}\label{nu}
 \omega_1 = \omega_1^0+\nu,\qquad   \omega_2 = \omega_2^0-\nu.
\end{equation}

\subsection{Simplified single-mode pair representation}

We start off with the theoretical description that can be found rather
ubiquitously in the literature (see for instance
\cite{Rubin94,TSuhara,TSuhara2007,Martin08,Martin10,KuklewiczPhD,Sciarrino}). As
we will see, although this description will give the correct predictions at the
optimal temperature of the ppKTP crystal, it will be insufficient at
other temperatures.

Since we are selecting only the pairs that have a wavelength of 810 nm
(with an interference filter with~1~nm bandwidth in our case), we can
replace Eq~\eqref{SPDC} by a single frequency mode, with two orthogonal
polarizations \cite{Rubin94,TSuhara,TSuhara2007}
\begin{equation}\label{HVstate}
    |\psi\rangle \propto a_\mathrm{in,H}^\dag(\omega_p/2)\; a_\mathrm{in,V}^\dag (\omega_p/2)|0\rangle=|\mathrm{H}\rangle|\mathrm{V}\rangle,
\end{equation}

In order to dispatch the photons to Alice and Bob, the pairs of
collinear photons produced in the ppKTP crystal are sent to a non
polarizing beam splitter (NPBS). For a photon impinging with a spatial
mode $|\mathrm{in}\rangle$ on an ideal 50/50 NPBS, the output state in
terms of the transmitted mode $|\mathrm{a}\rangle$ and the reflected
mode $|\mathrm{b}\rangle$ (sent respectively to Alice and Bob; see
Fig.~\ref{RTfig}) depends on the initial polarization of the photon
\cite{TSuhara}:
\begin{equation}\label{BStransfo}
\begin{aligned}
    |\mathrm{H}\rangle|\mathrm{in}\rangle &   \xrightarrow{\mathrm{NPBS}} \frac{1}{\sqrt{2}} \big(|\mathrm{H}\rangle|\mathrm{a}\rangle+i|\mathrm{H}\rangle|\mathrm{b}\rangle \big)
    \\
     |\mathrm{V}\rangle|\mathrm{in}\rangle &   \xrightarrow{\mathrm{NPBS}} \frac{1}{\sqrt{2}} \big(|\mathrm{V}\rangle|\mathrm{a}\rangle-i|\mathrm{V}\rangle|\mathrm{b}\rangle \big)
\end{aligned}
\end{equation}
where the origin of the minus sign is due to the phase shift $\pi$ for a reflected wave with horizontal polarization at a beam splitter.

If we now consider two orthogonally polarized photons impinging on the
beam splitter with the same input spatial mode, we can write, using
the spatial mode as a shorthand index for the polarization mode:
\begin{eqnarray}\label{HVBS}\nonumber
    |\mathrm{H}\rangle_\mathrm{in}|\mathrm{V}\rangle_\mathrm{in} \xrightarrow{\mathrm{NPBS}}
\frac{1}{2} &\big(&
    |\mathrm{H}\rangle_\mathrm{a}|\mathrm{V}\rangle_\mathrm{a}
    -i|\mathrm{H}\rangle_\mathrm{a}|\mathrm{V}\rangle_\mathrm{b}\\
    & &+i|\mathrm{H}\rangle_\mathrm{b}|\mathrm{V}\rangle_\mathrm{a}
    +|\mathrm{H}\rangle_\mathrm{b}|\mathrm{V}\rangle_\mathrm{b}
    \big).
\end{eqnarray}

The usual argument at this point
\cite{Rubin94,TSuhara,TSuhara2007,Martin08,Martin10,KuklewiczPhD,Sciarrino}
is that the cases in which the two photons exit through the same port
(that is, $|\mathrm{H}\rangle_\mathrm{a}|\mathrm{V}\rangle_\mathrm{a}$
and $|\mathrm{H}\rangle_\mathrm{b}|\mathrm{V}\rangle_\mathrm{b}$) can
be discarded because of the postselection of the photons. Only those
pairs with one photon for Alice and one photon for Bob are of interest
for the experimental results.

After making the substitution
$|\mathrm{H}\rangle_\mathrm{b}|\mathrm{V}\rangle_\mathrm{a}\rightarrow|\mathrm{V}\rangle_\mathrm{a}|\mathrm{H}\rangle_\mathrm{b}$,
and renormalizing, the state of the pairs of photons detected in
coincidence by Alice and Bob can be written as the singlet state:
\begin{equation}\label{singlet}
    |\Psi^-\rangle_\mathrm{ab} = \frac{1}{\sqrt{2}} \big[|\mathrm{H}\rangle_\mathrm{a}|\mathrm{V}\rangle_\mathrm{b}-|\mathrm{V}\rangle_\mathrm{a}|\mathrm{H}\rangle_\mathrm{b} \big],
\end{equation}
which is a polarization-entangled state.

\begin{figure}
\includegraphics[width=8.75cm]{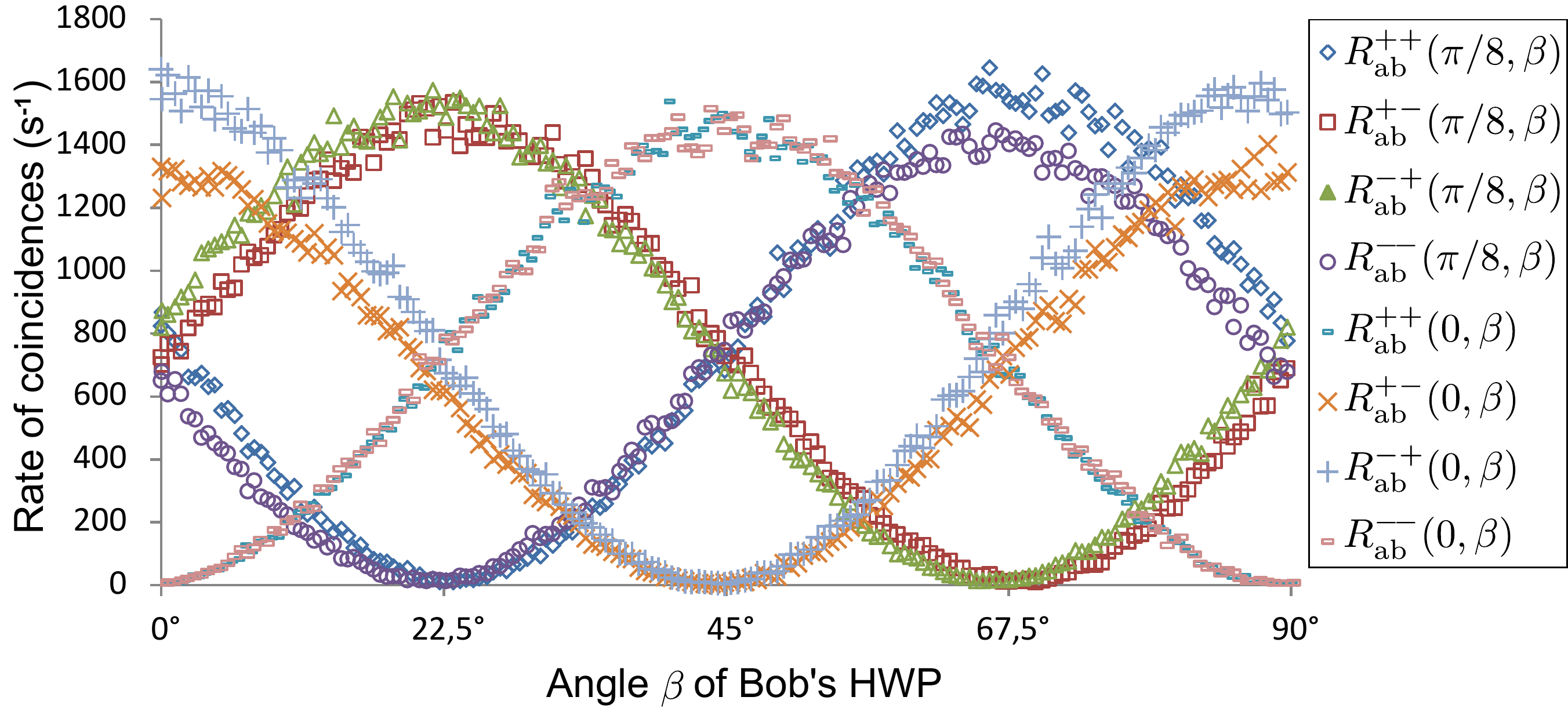}
\caption{\label{Coincidences} Rates of coincidences in fixed bases
  while the temperature of the crystal is maintained at the optimum
  temperature of $35.1^\circ$C. Alice keeps her measurement setting
  fixed, at $\alpha=0$ (rectilinear basis) or $\alpha=\pi/8$ (diagonal
  basis), while Bob's HWP setting $\beta$ is varied from 0 to 90 degrees. The
  coincidences exhibit a visibility of 99.6\% in the rectilinear basis
  and of 98.5\% in the diagonal basis (without substraction of
  accidental), which would amount to a CHSH function $S\simeq2.80$.}
\end{figure}

Starting from this post-selected state, we want to calculate the
predictions for the coincidence counting rates. The rotation imparted
to the polarization of a photon by a half wave plate oriented with an
angle $\theta$ with respect to the horizontal can be written as:
\begin{equation}\label{HWPtransfo}
\begin{aligned}
    |\mathrm{H}\rangle  &\xrightarrow{\mathrm{HWP}\;\theta}& &\cos2\theta|H\rangle+\sin2\theta|V\rangle&
    \\
     |\mathrm{V}\rangle  &\xrightarrow{\mathrm{HWP}\;\theta}&  -&\sin2\theta|H\rangle +\cos2\theta|V\rangle&.
\end{aligned}
\end{equation}
Using the transformations for the half-wave plate of
Eqs.~(\ref{HWPtransfo}), with $\theta=\alpha$ for the photon going to
Alice and $\theta=\beta$ for the photon going to Bob, the rates of
coincidences measured by Alice and Bob take the simple and well-known form
associated to the singlet state:
\begin{equation}\label{expoptC}
\begin{aligned}
    R^{++}_\mathrm{ab} = R^{--}_\mathrm{ab} &\varpropto \frac{1}{2} \sin^2 2(\alpha-\beta)\\
    R^{+-}_\mathrm{ab} = R^{-+}_\mathrm{ab} &\varpropto \frac{1}{2} \cos^2 2(\alpha-\beta).
\end{aligned}
\end{equation}
With good approximation the rate of coincidences that we have  measured
at the optimal temperature are indeed of this form (see
Fig.~\ref{Coincidences}). The visibility (or contrast) of
the coincidences was indeed slightly less than ideal, with a visibility
of 99.6\% in the rectilinear basis and of 98.5\% in
the diagonal basis (without substraction of accidental).

In the above theory, the coincidence rates depend on the polarizers angles, but are independent of
the temperature. In particular, in
the diagonal basis, $\alpha=\beta=\pi/8$, we predict
$E_{ab}(\frac{\pi}{8},\frac{\pi}{8}) = -1 $ at all temperatures. As we
see in Fig.~\ref{CorrelationNonOptTemp} this is true only for the
optimal temperature, and approximately true for a few other
temperatures.  Of particular interest are the temperature values for
which the correlation becomes positive and close to 1, as pointed out in detail above.  We conclude that
the single-mode pair representation fails at temperatures different
from the optimal, and we proceed to give a more detailed analysis.

\subsection{Full theory using the frequency-entangled state}

We apply the conventional description (see for example Ref. \cite{Shih}) for the propagation of the downconverted state, Eq \eqref{SPDC}, through the optical elements. The details are given in Appendix~\ref{setup}. Here we only describe how to account for the effect of the interference filter and give the results for the two-photon amplitude.

When placing the interference filter in the beam we have to modify the
two photon state of Eq \eqref{SPDC}

\begin{equation}\label{TPSGF}
  \ket{\psi} = W\int d\nu
  f(\nu)F(\nu)
  a_{in,H}^\dag(\omega_1) a_{in,V}^\dag(\omega_2)\ket{0}
\end{equation}
with the filter function $F(\nu)=\bar{G}(\omega_1)\bar{G}(\omega_2)$.
The function $\bar{G}(\omega)$ describes the amplitude for a photon of
frequency $\omega$ passing the filter. We assume that the function
$\bar G$ has a peak centered on $\omega_p/2$, so that we can write
$\bar{G}(\omega) = G(\omega-\omega_p/2)$ with the function $G(\omega)$
having a peak centered at zero. Defining 
$\mu = \omega_1^0-\omega_2^0$, Eq. \eqref{nu} gives
\begin{equation}\label{om}
 \omega_1 = \frac{\omega_p}{2} +  \frac{\mu}{2} + \nu \qquad\qquad 
 \omega_2 = \frac{\omega_p}{2} - \frac{\mu}{2} - \nu 
\end{equation}
and we get 
\begin{equation}\label{FGG}
 F(\nu) = G\left(\frac{\mu}{2}+\nu\right)G\left(\frac{-\mu}{2}-\nu\right).
\end{equation}
We will restrict ourselves to symmetric filter functions, $G(-\nu) = G(\nu)$, and  we have that 

\begin{equation}\label{FG2}
 F(\nu) = G(\nu+\frac{\mu}{2})^2 .
\end{equation}

To calculate the coincidence rates we define the amplitude $\psi_{++}(t_A,t_B)$ for the  detection of one photon in detector $A+$ at time $t_A$ and one in $B+$ at time $t_B$:

\begin{equation}
  \psi_{++}(t_A,t_B) = \langle0| E_{A+}^{(+)}E_{B+}^{(+)} |\psi\rangle
\end{equation}
with similar definitions for $\psi_{+-}$, $\psi_{-+}$ and $\psi_{--}$.
Here $E_{A+}^{(+)}$ is the annihilation part of the field operator for
the electric field at detector $A+$, and similar for the other
detectors. The exact form of these operators are given in
Appendix~\ref{setup}, where it is shown that when $\alpha=\beta=\pi/8$
\begin{equation}\label{psiAB}
\begin{aligned}
  \psi_{++}(t_B,t_A) &=   -\frac{W}{4}\int d\nu f(\nu)F(\nu)\\
   &\left[e^{-i\omega_1\tau_{A+}^H-i\omega_2\tau_{B+}^V}
     -e^{-i\omega_2\tau_{A+}^V-i\omega_1\tau_{B+}^H}\right].
\end{aligned}
\end{equation}
where
\begin{equation}\label{tau}
\begin{aligned}
  \tau_{M\pm}^H &= t_{M\pm}-z_{M\pm}/c-\tau_c, \\
  \tau_{M\pm}^V &= t_{M\pm}-z_{M\pm}/c.
\end{aligned}
\end{equation}
is the time a photon  exits the ppKTP if it is detected at the
detector $M\pm$ at time $t_{M\pm}$ ($M=A, B$ labels the
detectors). $z_{M\pm}$ is the distance from the end of the ppKTP
crystal to detector $M\pm$ and the time delay $\tau_c = L_cD$ (where
$L_c$ is the length of the compensating crystal) is the difference in
the time taken by photons with different polarisations to pass the
compensating crystal. 

The coincidence rate is then given by the standard expression (see
e.g. \cite{Shih})

\begin{equation}\label{rcint}
  R^{++}_\mathrm{ab} = \frac{1}{2T}\int_0^{2T} dt_+ \int dt_- |\psi_{++}|^2
\end{equation}
where $t_\pm = t_A\pm t_B$.
In the appendix we show that when  $\alpha=\beta=\pi/8$ this gives 

\begin{equation}\label{Rpp}
 R_\mathrm{ab}^{++} = R_\mathrm{ab}^{--} =  R_0\left[I_1-I_2\right]
\end{equation}
\begin{equation}\label{Rpm}
 R_\mathrm{ab}^{+-} = R_\mathrm{ab}^{-+} =  R_0\left[I_1+I_2\right]
\end{equation}
with
\begin{equation}\label{I12}
\begin{aligned}
 I_1 &= 4\int d\zeta\frac{G(\zeta+m/2)^4}{\zeta^2}\sin^2\frac{\zeta}{2}\\
 I_2 &= 4\int d\zeta\frac{G(\zeta+m/2)^4}{\zeta(\zeta+m)}\sin\frac{\zeta}{2}\sin\frac{\zeta+m}{2}
\end{aligned}
\end{equation}
where we use the dimensionless variables 
\begin{equation}\label{m}
 \zeta = \nu DL,  \qquad m = \mu DL,
\end{equation}
where we write $G(\omega)$ instead of $G(\omega/DL)$, and  where we took the
length $L_c$ of the compensating crystal to be half the length $L$ of the
ppKTP crystal, as the case in the experiment.

According to Eq.~\eqref{correlation}  the correlation in the diagonal basis is

\begin{equation}
 E_\mathrm{ab} (\pi/8,\pi/8) = -\frac{I_2}{I_1}.
\end{equation}
The integrals $I_1$ and $I_2$ are probably difficult to calculate for
most filter functions, and we are satisfied with numerical solutions of
these integrals.

\subsection{Comparing with the experimental data}

To compare the theory with experiments, we need two things: the
bandwidth of the interference filter and the scaling of the frequency difference between the two photons
$\mu(T)$ with the temperature $T$.

We assume a Gaussian filter 
\begin{equation}\label{GaussFilter}
 G(\zeta) = e^{-(\zeta/Z)^2}
\end{equation}
In the experiments we used filters with nominal bandwidths of 1 and 10
nm. For the 1 nm filter we measured the absorption of the filter
spectroscopically and fitted a Gaussian of the form
$e^{-\left(\frac{\lambda-\lambda_0}{W}\right)^2}$ with $\lambda_0 =
810$ nm. The average of two measurements gave a best fit $W= 0.64$ nm
which is not far from the nominal specification. This has to be translated to the
frequency-domain filter function \eqref{GaussFilter}. Since $G(\zeta)$ is
the amplitude for the photon to pass the filter, the probability is
$G(\zeta)^2$. Thus we set
$2(\zeta/Z)^2=\left(\frac{\lambda-\lambda_0}{W}\right)^2$. 
Recalling that $\zeta = \Delta\omega DL$, where $ \Delta\omega$ is the difference between the photon frequency and the center frequency of the filter, we have that 

\begin{equation}
 \zeta = 2\pi DLc\left(\frac{1}{\lambda_0}-\frac{1}{\lambda}\right)\approx
  2\pi DLc\frac{\lambda-\lambda_0}{\lambda_0^2}
\end{equation}
where the approximation is valid as long as $\lambda-\lambda_0\ll\lambda_0$ which is appropriate for a narrow filter. Using that $D = 1/c_2-1/c_1 = (n_2-n_1)/c$ where $n_1 = 1.75$
and $n_2 = 1.84$ are the indices of refraction of the two beams at the
optimal temperature we get 

\begin{equation}
 Z = \frac{2\sqrt{2}\pi(n_2-n_1)L}{\lambda_0^2}W = 7.8
\end{equation}
where $L=10$ mm is the length of the ppKTP crystal. For the 10 nm
filter we do not have similar data, but we can assume that it has an
$A=80$, about 10 times as large as the 1 nm filter. It is anyway so
large that there will be no significant difference between the
predicted result with this filter and with no filter at all.

To fully predict the experimental results, we also have to know the
temperature dependence $\mu(T)$ of the difference in the center
frequencies of the two downconverted beams. This can in principle be
found from the phase matching conditions \eqref{omegaSPDC} and
\eqref{kSPDC}, using the temperature dependence of the indices of
refraction and the thermal expansion of the poling period. We tried
several sets of published Sellmeier coefficients for KTP
\cite{Kato,Emanueli,Zhao}, and they give substantially different
results.  None of them predict exactly the correct optimal temperature
or a $\mu(T)$ that fits the results accurately. The optimal
temperature is known from the data of
Fig.~\ref{CoincidencesNonOptTemp}, and differs from the predicted
value by a few degrees for the best sets of Sellmeier
coefficients. The value of $\mu(T)$ is typically off by 10-50\%. We
note that a similar situation is reported in Ref.~\cite{fedrizzi}. It
seems likely that the exact temperature dependence of the refraction
indices is slightly different for different crystals, depending on the
purity of the crystal and the growth and poling conditions. We have
therefore used the measured value of the optimal temperature, and
assumed a linear dependence $\mu(T) = a(T-T_\mathrm{opt})$ with the
propotionality constant $a$ as a fitting parameter.

The results of the numerical evaluation of the integrals $I_1$ and
$I_2$ with the corresponding filter functions gives the correlation
functions shown as lines in Fig. \ref{CorrelationNonOptTemp}. The only
fitting parameter is the propotionality constant $a$, and the same
scaling is used for both filters. We see that the theory fully
accounts for the experimental result. Observe that even with a 1 nm
filter the predictions are different from the simple single-frequency
two-mode description discussed above. This means that the bandwidth of
this filter is still too large for the simplified single-frequency two-mode description to be sufficient. It seems that the simplified description gives a satisfactory description only
at the optimal temperature.
 We will
 now investigate why it works at this temperature, and
how narrow the filters must be so that the results are the same as with a single-frequency mode.

\section{Why the two-mode description works at the optimal temperature}

\subsection{Without the interference filter}\label{sec:wavefronts}

Let us go back to Eq \eqref{psiAB} and insert the definition of
$f(\nu)$. If we  consider the case where there is no
interference filter, so that $F(\nu)=1$ we get (for simplicity, we
write $\tau_A^H$ instead of $\tau_{A+}^H$ and similarly for all other
quantities, it being understood that we consider the $++$ correlation
in these formulas)

\begin{equation}\label{intznu}
\begin{aligned}
  \psi(t_B,t_A) \sim & \int_{-L}^0dz\int d\nu e^{iDz\nu}\\
  &   \left[e^{-i\omega_1\tau_A^H-i\omega_2\tau_B^V}
     -e^{-i\omega_2\tau_A^V-i\omega_1\tau_B^H} \right]
\end{aligned}
\end{equation} 
Here, $e^{-i\omega_1\tau_A^H}$ is the phase change of a horizontally
polarized wave with frequency $\omega_1$ and wave velocity $c$
propagating from the origin (end of the ppKTP) to detector A+, and similiarly for the other phases. The
factor $e^{iDz\nu}$ is due to the phase changes of two waves
with frequencies $\omega_1$ and $\omega_2$ and wave velocities $c_1$
and $c_2$ from the point $z$ to the origin, multiplied by the phase
$e^{ik_pz}$ of the pump field at point $z$. The interpretation of
Eq \eqref{intznu} is then that it is a sum of waves starting at all
points $-L<z<0$ inside the ppKTP crystal and with all possible
frequencies of the two outgoing waves, while keeping the energy
conversion equation $\omega_1+\omega_2=\omega_p$. This is in the
spirit of Feynman's path integral approach to quantum mechanics, where
the total amplitude of a process is a sum over the amplitudes for all
possible ways that the process can take place in.

 We can give a more detailed geometric interpretation of this
 expression. Consider the first term in Eq.~\eqref{intznu}, which, as
 is well known~\cite{Rubin94,Shih}, gives

\begin{equation}\label{firstTerm}
\int_{-L}^0dz\int d\nu e^{iDz\nu}
   e^{-i\omega_1\tau_A^H-i\omega_2\tau_B^V} = \Pi(\tau_-/D)e^{-i\omega_1^0\tau_A^H-i\omega_2^0\tau_B^V}
\end{equation}
where $\Pi(x) = 1$ if $-L\leq x\leq0$ and zero otherwise. 

\begin{figure}
\includegraphics[width=\linewidth]{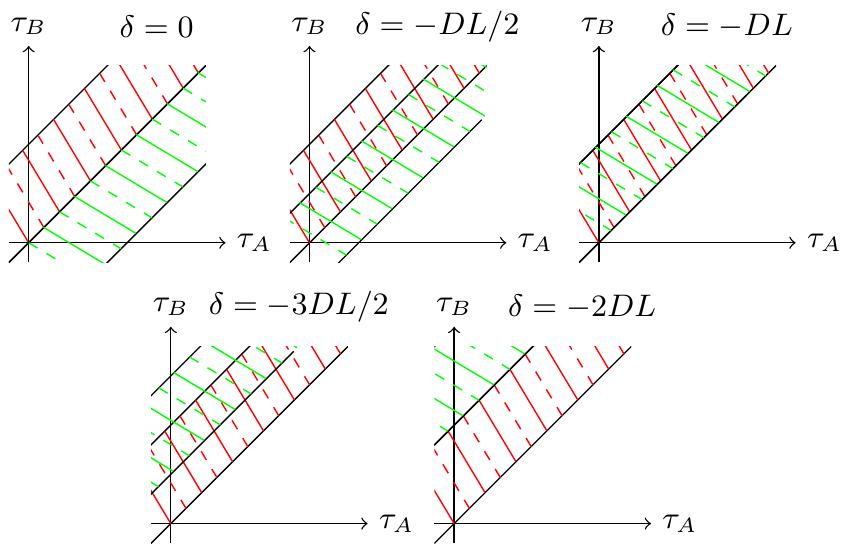}
\caption{\label{wavefrontsDelta}Lines of equal phase for different thickness of the compensating crystal, represented by the different values of $\delta -2\tau_c$. The compensating crystal used in the experiment corresponds to $\delta = -DL$, which leads to maximal overlap  of the wave bands. The lines of equal phase for the first term in Eq.~\eqref{intznu} are shown in red, while those for the second term are green. To clarify the phase relations, we have shown a set of solid lines where the phase difference between each solid line is $2\pi$. The dashed lines interspersed between the solid lines represent a phase shift of $\pi$ with respect to the solid lines.  Outside of the two bands, the wavefuction is zero, as seen from  Eq.~\eqref{firstTerm}, and two-photon interference only occurs when the two bands overlap.  }
\end{figure}

Recalling Eq.~\eqref{tau} we see that if we consider a fixed time,
which we can choose as  $t=0$, then $\tau_M^P$ (for detector $M=A,B$ and polarization
$P=H,V$) describes the changing phase of the wave in space along the
given path. To get a full picture, we can plot the waves along the two
paths along two orthognal axes using the times $\tau_M^P$ as
coordinates, and Eq.~\eqref{firstTerm} describes a set of plane waves in this abstract space.
As seen on the right hand side of Eq.~\eqref{firstTerm}, the waves
will interfere destructively outside of a certain band in the
$(\tau_A,\tau_B)$-plane. Inside this band, we have a plane wave with
the direction of the lines of equal phase (wavefronts) determined by the center frequencies
$\omega_1^0$ and $\omega_2^0$.  In the second term of
Eq.~\eqref{intznu} the detectors $A$ and $B$ are exchanged, so the
image is reflected in the diagonal of the $(\tau_A,\tau_B)$-plane. In
addition, there is a shift for finite $\delta = -2\tau_c$ (where $\tau_c = L_cD$ is the difference between the times the beams with orthogonal polarization uses in passing the compensating crystal
of length $L_c$):

\begin{equation}\label{33}
\begin{aligned}
\tau_A^V &= \tau_A^H-\delta/2\\
\tau_B^H &= \tau_B^V+\delta/2
\end{aligned}
\end{equation}
which gives the picture shown in Fig.~\ref{wavefrontsDelta}, where
the lines of equal phase from the first term of Eq.~\eqref{intznu} are shown in
red, while the lines of equal phase from the second term are green.
 Our experiments were performed with a compensating crystal whose
 length was half of the ppKTP crystal, which gives $\delta=-DL$,
 corresponding to maximal overlap of the wave bands, and therefore
 maximal interference between the two-photon wavepackets. From now on,
 we will only consider this value of $\delta$.

\begin{figure}
\includegraphics[width=\linewidth]{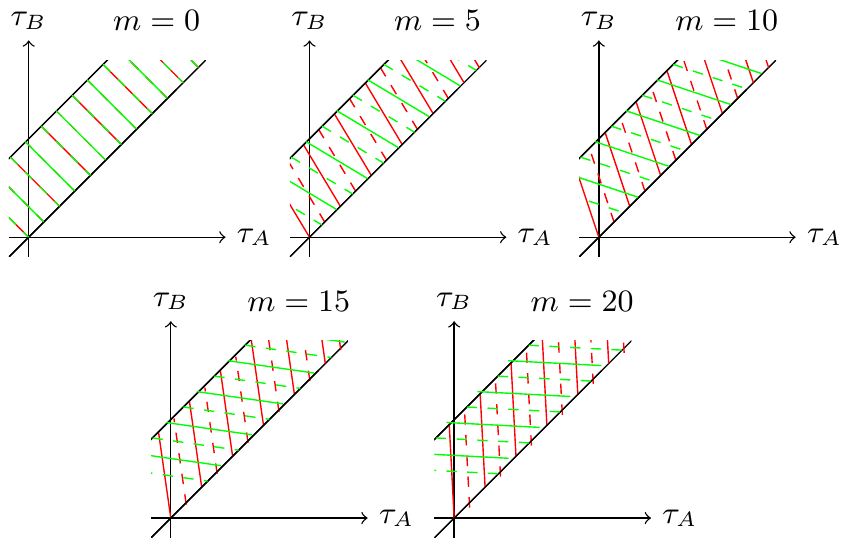}
\caption{\label{wavefrontsMu}Lines of equal phase for different temperatures, as represented by the value  of $m = \mu(T)DL$, where $\mu(T) = \omega_1^0-\omega_2^0$ is the temperature dependent difference between the center frequencies of the two beams. The case $m=0$ corresponds to the optimal temperature,
while larger values of $m$ correspond to increasing deviations from the optimal temperature.
 All figures are for a compensating crystal with a length half that of the ppKTP as used in the experiments, which means that $\delta=-DL$ and that the bands from the two terms of Eq.~\eqref{intznu} overlap fully, giving maximal two-photon interference. Colors are the same as in Figure \ref{wavefrontsDelta}. As we see, for $m=0$, the two waves always meet in phase opposition, resulting in complete destructive interference at the center of the HOM dip.  }
\end{figure}

Consider now how this picture changes at different temperatures. As
the temperature deviates from the optimal, the center frequencies of
the two outcoming beams start to differ, and the slope of the
lines of equal phase changes. This is shown in Fig.~\ref{wavefrontsMu} in
terms of the dimensionless frequency difference $m$ as defined in
Eq.~\eqref{m}.
 For $m=0$ the lines of equal phase for the two terms are parallel,
 and since there is a relative minus between the two terms, we get
 complete destructive interference, corresponding to the center of the
 HOM dip. This is the origin of the perfect anticorrelation, $E=-1$,
 observed at the optimal temperature. As the temperature deviates from
 the optimal, the lines of equal phase get tilted, and we get the
 oscillations in the correlation observed in
 Fig. \ref{CorrelationNonOptTemp} with the 10 nm IF (red points),
 which we check numerically is virtually indistiguishable from no
 IF. To understand how the oscillations of the correlation arise from the
 tilting of the lines of equal phase, we go back to Eq.~\eqref{rcint}
 that gives the coincidence counting rate as an integral over the
 square of the two-photon amplitude (where we can change the
 integration variables to $\tau_\pm = \tau_A\pm\tau_B$). The integrand
 will be independent of $\tau_+$, so that we can fix $\tau_+$ at any value
 when considering the integral over $\tau_-$, which means integrating
 along a line crossing the wave band of Figure~\ref{wavefrontsMu}.
 For $m=0$ the lines of equal phase for the two terms always meet in
 phase opposition, while if we look at $m=5$ wee se that at the band edges
 the two terms meet in phase. However, to get the full picture, we
 have to remember that the wavefunction is a complex number, with real
 and imaginary parts. The coincidence rate is given by the integral in
 Eq.~\eqref{rcint}, and it is not so easy to visualize this complex behaviour using the
 diagrams of Fig.~\ref{wavefrontsMu}.  
We will now see how this wave pattern is
 changed by the insertion of the IF, which limits the frequencies to a
 band around half the pump frequency.

\subsection{The effect of the interference filter}\label{sec:filter}

\begin{figure}
\includegraphics[width=\linewidth]{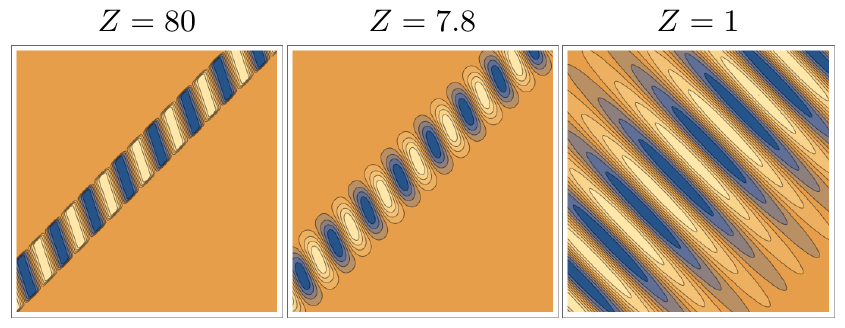}
\caption{\label{filter}Contour plots of the real part of the first
  term of Eq.~\eqref{intznuF} for different filter widths. The plots
  for $Z=80$ and $Z=7.8$ correspond to the experimental filters of 10
  nm and 1 nm, respectively. The plot for $Z=1$ corresponds to an even
  narrower filter than those used experimentally, and is added to show
  the effect of the interference filter more clearly. As without the
  filter, the graphs for the second term of Eq.~\eqref{intznuF} are
  similar, just reflected in the diagonal and shifted according to the
  value of $\delta$.}
\end{figure}

 With a filter in the setup we have to go back to \eqref{TPSGF}
and modify \eqref{intznu} accordingly (letting $\alpha=\beta=\pi/8$ as before): 

\begin{equation}\label{intznuF}
\begin{aligned}
  \psi(t_B,t_A) \sim & \int_{-L}^0dz\int d\nu F(\nu) e^{iDz\nu}\\
  &   \left[e^{-i\omega_1\tau_A^H-i\omega_2\tau_B^V}
     -e^{-i\omega_2\tau_A^V-i\omega_1\tau_B^H} \right]
\end{aligned}
\end{equation} 
As in \eqref{GaussFilter} we choose a Gaussian filter function. The
two-photon amplitude $\psi(t_B,t_A)$ is complex, so to visualize it
we use contour plots of the real part as functions over the
$(\tau_A,\tau_B)$-plane. In Fig.~\ref{filter} we show the first term of
Eq.~\eqref{intznuF} for different values of $Z$ and with $m=5$.  As we
decrease the filter bandwidth $Z$, two changes take place. First, the
band where coincidences can occur widens and becomes less sharpely
defined. This is natural since all frequencies are needed to provide
the sharp edges of the coincidence band (since  a sharp step in  a function only can be represented by  a Fourier transform when all frequencies are included). Second, the lines of equal phase  become
more aligned  45$^\circ$ to the axes.

We can now understand why the simple and popular single-frequency
two-mode description works at the optimal temperature, while it fails
at other temperatures. At the optimal temperature, the lines of equal phase are
parallel even without a filter. Adding a filter will smear out the
band where the amplitude is non-zero, but it will not change the
orientation of the lines of equal phase. So the interference effects between the
two terms in Eq.~\eqref{intznuF} are unchanged as the filter width is
changed. Only the overall amplitude, which is normalized away in the
correlation function, is sensitive to the filter. Therefore the
results of the experiments are the same as if only one frequency was
present, even if in the real experiment a superposition of frequencies
occur also at the optimal temperature. At temperatures different from
the optimal temperature, the lines of equal phase are not parallel without the filter, and
they are gradually made more parallel as the filter width is
decreased. The experiments are then sensitive to the frequency spread
of the downconverted photon beam, and the result depends on the filter
width.

\subsection{How narrow filter do we need to have a single-frequency description?}

In the limit of a very narrow filter, we would still expect to recover
the results of the one-frequency description. We can estimate how
narrow filter is needed at different temperatures using
Eq.~\eqref{I12}.  If $G(\zeta)$ is a sharply peaked function around zero,
we see that we will have contributions only for $\zeta$ close to
$-\frac{m}{2}$.  Then $\frac{\zeta}{2} = \frac{\zeta+m}{2} = \frac{m}{4}$, and
both integrals will be close to zero when $m= 4\pi n$. We are
interested in $m$ close to these points. We let $\zeta=-m/2 + \xi$ and
expand for small $\xi$:

\begin{equation}
\begin{aligned}
 f_1 &= \frac{\sin^2\frac{\zeta}{2}}{\zeta^2} = \frac{4\sin^2m/2}{m^2} - (\cdots)\xi \\
   &\quad+ \frac{m^2\cos m/2+ 48\sin^2m/4 - 8m\sin m/2}{m^4} \xi^2 \\
 f_2 &= \frac{\sin\frac{\zeta}{2}\sin\frac{\zeta+m}{2}}{\zeta(\zeta+m)} = \frac{4\sin^2m/2}{m^2} - \frac{m^2-16\sin^2m/4}{m^4} \xi^2
\end{aligned}
\end{equation}
where the linear term in $f_1$ is unimportant since it will disappear
when integrated over a symmetric filter function $G(\zeta)$. When $m$ is
close to $4\pi n$ (and $n\neq0$) we can ignore the variations of the
quadratic coefficients with $m$ and set $m= 4\pi n$ in the
trigonometric functions to get
\begin{equation}
\begin{aligned}
 f_1 &= \frac{4\sin^2\Delta m/2}{m^2} + \frac{\xi^2}{m^2}\\
 f_2 &= \frac{4\sin^2\Delta m/2}{m^2} - \frac{\xi^2}{m^2}
\end{aligned}
\end{equation}
where $\Delta m = m-4\pi n$.
For simplicity, assume a sharp filter function $G(\zeta)=1$ when $|\zeta|<Z$ and $G(\zeta)=0$ when $|\zeta|>Z$. Then 

\begin{equation}
 I_{1,2} = \frac{8Z}{3\Delta m^2}\left(12\sin^2\frac{\Delta m}{4}\pm Z^2\right)
\end{equation}
where the upper sign is for $I_1$ and the lower sign is for $I_2$.  We
see that if $Z>3\sqrt{2}\sin\frac{\Delta m}{4}\approx\Delta m$, the
second term dominates, and $I_2=-I_1$ which gives a correlation
$E_{AB} = -\frac{I_2}{I_1} = +1$, meaning that the outcome at Alice's arm is equal to that at Bob's arm. If $Z<\Delta m$ we have
$I_2=I_1$ which gives a correlation $E_{AB} = -1$, meaning that the outcome at Alice's arm is opposite to that at Bob's arm. Looking back to the experiments in Sec
\ref{exp} we see that the introduction of a narrower filter increased
the amplitude of the oscillations of the correlation, and made the
correlation closer to $+1$. At the same time, we now see that a
certain spread in frequency is necessary for the change between a correlation -1 and +1, and that with
 a sufficiently narrow filter, we expect the corelation to go
back to $-1$. For $m$ close to $4\pi n$ we then expect the
correlation to depend nonmonotonically on the filter width $Z$.

\section{Summary and discussion}

We have shown the results of a two-photon interference experiment where
entangled pairs of photons are created by parametric downconversion in
a ppKTP crystal. The downconversion process is most efficient at a
certain optimal temperature, and we have shown the effect of changing the
temperature away from this optimum. In previous
experiments~\cite{fedrizzi} it had been shown that the temperature dependence of the results reveals the
frequency entanglement of the two-photon state, as described by
Eq.~\eqref{SPDC}. This had been observed through the presence of a slightly positive value of the
correlation function~\eqref{correlation} between the two-photon
detections. 

We have demonstrated the same effect in a different setup, and we have found dramatic oscillations of the correlation between -1 and +1 as the temperature was changed. This happened in an experimental system where the
two photon beams are never recombined in a beamsplitter (after the primary non-polarizing beam splitter), but where the two
polarization states are mixed using half-wave plates. In addition we have
investigated the effect of inserting an interference filter before the
two beams are separated by a beamsplitter, to reduce the frequency
window over which the photons are spread. 

The most natural expectation
is then that the presence of a filter will reduce the effect of the
frequency entanglement, and thereby also the positive value of
Eq.~\eqref{correlation}. This is also the result of a simplified
theoretical description which ignores the frequency spread, and
includes only a single frequency mode. 
We observe the opposite, in the sense that the
correlation comes close to the maximal value of $+1$ at certain
temperatures, as seen in Fig.~\ref{CorrelationNonOptTemp}.

We have
provided a detailed theoretical analysis, which shows excellent
agreement with the experiments. We have also given an interpretation of the
analytical formulas in terms of the representation of the two-photon
amplitude in the space spanned by the detection times of the two
photons.

Our work demonstrates that the simplified single-mode pair
representation is insufficient to describe our experimental results.
The biphoton description involves a much more complicated
interplay between the two original photons created 'simultaneously' in
the non-linear crystal. The interplay involves in a very detailed
manner both frequency relations between the two original photons and
timing differences (phase shifts) on their way from where they are
created until the horizontal and vertically polarized contributions
have passed through the rest of the non-linear crystal, as well as through the
compensating crystal, the filters and the mixing NPBS. All these
details are contained in the biphoton state.

\appendix

\section{Details of the calculation of the two photon amplitude and coincidence rate}\label{setup}

We let operators $a_{a,H}$ and similar represent the field after the
NPBS but before the HWP and $a_{A,+}$ and similar after the HWP. The
effect of the HWP is then described by the transformation

\begin{equation}
\begin{aligned}
a_{A,+} &= \cos2\alpha\,a_{a,H} + \sin2\alpha\,a_{a,V}\\ 
a_{A,-} &= -\sin2\alpha\, a_{a,H} + \cos2\alpha\, a_{a,V} \\
a_{B,+} &= \cos2\beta\, a_{b,H} + \sin2\beta\, a_{b,V} \\
a_{B,-} &= -\sin2\beta\, a_{b,H} + \cos2\beta\, a_{b,V} 
\end{aligned}
\end{equation}
The effect of the PBS is to send the two different polarizations to
the two different detector, so that the field at one of the detectors
is

\begin{equation}
 E_{A,+}^{(+)} = i\int d\omega_AA(\omega_A)
    e^{-i\omega_A\tau_{A+}^P}a_{A,+}(\omega_A) 
\end{equation}
and similarly for the others. Here the time $\tau_{A+}^P$ depends on
the initial polarization $P£$ of the photon hitting the detector as will be
specified below. The action of the NPBS is to split each photon into
the two arms, with extra phase factors in some cases according to

\begin{equation}\label{NPBS}
\begin{aligned}
 a_{in,H}&\overset{\text{NPBS}}{\longrightarrow}
   \frac{1}{\sqrt{2}}\left(a_{a,H}+ia_{b,H}\right) \\
 a_{in,V}&\overset{\text{NPBS}}{\longrightarrow}
   \frac{1}{\sqrt{2}}\left(a_{a,V}-ia_{b,V}\right)
\end{aligned}
\end{equation}

 This gives the field operators at the detectors in
terms of the operators at the exit of the ppKTP:
\begin{widetext}
\begin{equation}
\begin{aligned}
 E_{A+}^{(+)} &= \frac{i}{\sqrt{2}}\int d\omega_AA(\omega_A)
    \left[e^{-i\omega_A\tau_{A+}^H}\cos2\alpha\,a_{in,H}
        + e^{-i\omega_A\tau_{A+}^V}\sin2\alpha\,a_{in,V}  \right]\\
 E_{A-}^{(+)} &= \frac{i}{\sqrt{2}}\int d\omega_AA(\omega_A)
    \left[-e^{-i\omega_A\tau_{A-}^H}\sin2\alpha\,a_{in,H}
        + e^{-i\omega_A\tau_{A-}^V}\cos2\alpha\,a_{in,V}  \right]\\
 E_{B+}^{(+)} &= \frac{1}{\sqrt{2}}\int d\omega_BA(\omega_B)
    \left[e^{-i\omega_B\tau_{B+}^H}\cos2\beta\,a_{in,H}
        - e^{-i\omega_B\tau_{B+}^V}\sin2\beta\,a_{in,V}  \right]\\
 E_{B-}^{(+)} &= \frac{1}{\sqrt{2}}\int d\omega_BA(\omega_B)
    \left[-e^{-i\omega_B\tau_{B-}^H}\sin2\beta\,a_{in,H}
        - e^{-i\omega_B\tau_{B-}^V}\cos2\beta\,a_{in,V}  \right]
\end{aligned}
\end{equation}
Here

\begin{equation}
  \tau_{M\pm}^H = t_{M\pm}-z_{M\pm}/c-\tau_c, \qquad  \tau_{M\pm}^V = t_{M\pm}-z_{M\pm}/c
\end{equation}
is the relative time a photon exited the ppKTP if it is detected at the
detector $M\pm$ at time $t_{M\pm}$ ($M=A, B$ labels the
detectors). The time delay $\tau_c = L_cD$ (where $L_c$ is the length
of the compensating crystal) is the difference in the time it takes
for photons with different polarisations to pass the compensating
crystal.

Using this (and referring to Eqs.~\eqref{fnu} and \eqref{TPSGF}) we get 
\begin{equation}\label{psiABApp}
\begin{aligned}
 \psi_{++}(t_A,t_B) &= \langle0| E_{A+}^{(+)}E_{B+}^{(+)} |\psi\rangle \\
  &= -\frac{W}{2}\int d\omega_A d\omega_B d\nu f(\nu)F(\nu)\bra{0}
   \left[e^{-i\omega_A\tau_{A+}^H}\cos2\alpha a_{in,H}(\omega_A) 
    + ie^{-i\omega_A\tau_{A+}^V}\sin2\alpha a_{in,V}(\omega_A) \right]\\
   &\qquad\qquad  \left[ie^{-i\omega_B\tau_{B+}^H}\cos2\beta a_{in,H}(\omega_B) 
    + e^{-i\omega_B\tau_{B+}^V}\sin2\beta a_{in,V}(\omega_B) \right]
    a_{in,H}^\dag(\omega_1) a_{in,V}^\dag(\omega_2)\ket{0}\\
  &= -\frac{W}{2}\int d\omega_A d\omega_B d\nu f(\nu)F(\nu)
   \left[e^{-i\omega_A\tau_{A+}^H-i\omega_B\tau_{B+}^V}\cos2\alpha\sin2\beta
      \delta(\omega_A-\omega_1)\delta(\omega_B-\omega_2)\right. \\
  & \left.\hspace{4cm} -e^{-i\omega_A\tau_{A+}^V-i\omega_B\tau_{B+}^H}\sin2\alpha\cos2\beta      \delta(\omega_A-\omega_2)\delta(\omega_B-\omega_1)\right]\\
 &=    -\frac{W}{2}\int d\nu f(\nu)F(\nu)
   \left[e^{-i\omega_1\tau_A^H-i\omega_2\tau_B^V}\cos2\alpha\sin2\beta
     -e^{-i\omega_2\tau_A^V-i\omega_1\tau_B^H}\sin2\alpha\cos2\beta \right]
\end{aligned}
\end{equation}
Similarly we find that 

\begin{equation}\label{psipm2}
 \psi_{+-}(t_A,t_B)
 =    -\frac{W}{2}\int d\nu f(\nu)F(\nu)
   \left[e^{-i\omega_1\tau_A^H-i\omega_2\tau_B^V}
         \cos2\alpha\cos2\beta
     +e^{-i\omega_2\tau_A^V-i\omega_1\tau_B^H}
         \sin2\alpha\sin2\beta \right]
\end{equation}

which for $\alpha=\beta=\pi/8$ (and we write the following equations only in this case) is the same as $\psi_{++}$ except for
the sign of the second term. The remaining $\psi_{--}$ and $\psi_{-+}$
are related to the ones given by simple symmetry relations.

Defining $\tau_\pm = \tau_A^H-\tau_B^V$ we can rewrite Eq. \eqref{psiABApp} for $\alpha=\beta=\pi/8$ as (see also Eqs.~\eqref{om}--\eqref{FG2} and \eqref{33} )
\begin{equation}
 \psi_{++}(t_A,t_B) =    -\frac{W}{2}\int d\nu f(\nu)F(\nu)e^{-i\frac{\omega_p}{2}\tau_+}
   \left[e^{-i(\frac{\mu}{2}+\nu)\tau_-}
     -e^{i(\frac{\mu}{2}+\nu)(\tau_--\delta)} \right]
\end{equation}

For the coincidence rate we  then find 

\begin{equation}
\begin{aligned}
R_\mathrm{ab}^{++} &= \frac{D^2R_0}{2}\int d\tau_-\int d\nu_1d\nu_2\int_{-L}^0dz_1dz_2 
  e^{iD(z_2\nu_2-z_1\nu_1)}\\
&\left[e^{i(\nu_1-\nu_2)\tau_-} +e^{i(\nu_2-\nu_1)(\tau_--\delta)}  
  - e^{i(\mu+\nu_1+\nu_2)\tau_--\frac{i}{2}\mu\delta-i\nu_2\delta}
  - e^{-i(\mu+\nu_1+\nu_2)\tau_-+\frac{i}{2}\mu\delta+i\nu_1\delta}
\right]F(\nu_1)F(\nu_2)\\
&=\frac{D^2R_0}{2}\int d\nu_1d\nu_2\int_{-L}^0dz_1dz_2 
  e^{iD(z_2\nu_2-z_1\nu_1)}\\
&\qquad\left[2\delta(\nu_1-\nu_2) - \delta(\mu+\nu_1+\nu_2)\left(
  e^{-\frac{i}{2}\mu\delta-i\nu_2\delta}
  + e^{\frac{i}{2}\mu\delta+i\nu_1\delta}\right)
\right]F(\nu_1)F(\nu_2)\\
&=D^2R_0\int d\nu\int_{-L}^0dz_1dz_2 
 \left[F(\nu)^2e^{iD(z_2-z_1)\nu} - F(\nu)F(-\mu-\nu)e^{iD(z_1+z_2)\nu+iDz_1\mu} \;
  e^{-\frac{i}{2}\mu\delta-i\nu\delta}
\right]\\
&=R_0\int d\nu
 \left[\frac{F(\nu)^2}{\nu^2}\left[1-e^{-iDL\nu}\right]\left[1-e^{iDL\nu}\right]
 + \frac{F(\nu)F(-\mu-\nu)}{\nu(\mu+\nu)}\left[1-e^{-iDL(\nu+\mu)}\right]\left[1-e^{-iDL\nu}\right]
  e^{-\frac{i}{2}\mu\delta-i\nu\delta}
\right]\\
\end{aligned}
\end{equation}

\end{widetext}
which gives Eqs.~\eqref{Rpp}--\eqref{I12} of the main text when the length $L_c$ of the compensating crystal is half the length $L$ of the ppKTP crystal.

\bibliographystyle{plain}

\end{document}